\documentclass[a4paper]{PoS}
\usepackage{amsmath,amssymb,amsthm,bm,bbm,color}
\usepackage{graphics,subfigure,rotating}

\title{Smearing Center Vortices}
\ShortTitle{Smearing Center Vortices}

\author{\speaker{Roman H\"ollwieser}$^1$$^2$ and Michael Engelhardt$^1$\\
E-mail: \email{hroman@kph.tuwien.ac.at}, \email{engel@nmsu.edu}\\
$^1$Department of Physics, New Mexico State University, PO Box 30001,
Las Cruces, NM 88003-8001, USA\\
$^2$Institute of Atomic and Subatomic Physics, Vienna University of
Technology, Operngasse 9, 1040 Vienna, Austria}

\abstract{
We smear Z(2) center vortices in lattice gauge configurations such as to recover
thick vortices with full SU(2) Yang-Mills information.
In particular, we address the problem that using Z(2)
configurations in conjunction with overlap  (or chirally improved) fermions is
problematic due to their lack of smoothness. Our method allows us to regain this
smoothness and simultaneously maintain the center vortex structure. We verify
our method with various gluonic and fermionic observables and find good
agreement between smeared vortex configurations and full SU(2).\thanks{The numerical simulations were performed at the Phoenix and Vienna Scientific Cluster (VSC) at VUT and the Riddler Cluster at NMSU. This research was supported by the Erwin Schr\"odinger Fellowship program of the Austrian Science Fund FWF (``Fonds zur F\"orderung der wissenschaftlichen Forschung'') under Contract No. J3425-N27 (R.H.) and the U.S. DOE through the grant DE-FG02-96ER40965 (M.E.).}
}

\FullConference{The 32nd International Symposium on Lattice Field Theory,\\
		23-28 June, 2014\\
		Columbia University New York, NY}

\begin{document}

\section{Introduction}

The vortex model~\cite{'tHooft:1977hy,Vinciarelli:1978kp,Cornwall:1979hz}
assumes that the center of the gauge group is crucial for confinement. The center
degrees of freedom can be extracted from gauge field configurations by maximal
center gauge (MCG) and center projection~\cite{DelDebbio:1996mh}. These d.o.f. are dubbed P-vortices and can be viewed as two-dimensional surfaces on the
four-dimensional lattice. They are thought to approximate objects already
present in configurations before the extraction step. These latter objects are
called thick vortices, carry quantized magnetic center flux and are responsible
for confinement according to the vortex model.  The extracted P-vortex surfaces
are complicated, unorientable random surfaces percolating through the lattice. These
and other P-vortex properties are in good agreement with the requirements to
explain confinement, which was shown both in lattice Yang-Mills theory and
within a corresponding infrared effective model, see {\it e.g.}~\cite{DelDebbio:1996mh,Langfeld:1997jx,Kovacs:1998xm,Engelhardt:1999wr,Engelhardt:2003wm,Hollwieser:2014lxa,Altarawneh:2014aa}.
The vortex model can be applied to other infrared features of QCD not
immediately related to confinement, such as the topological properties of gauge
fields. In particular it was shown how the topological susceptibility present in
QCD can be calculated from the extracted P-vortex surfaces~\cite{
Bertle:2001xd,Engelhardt:2000wc,Engelhardt:2010ft} and vortices are also able to explain
chiral symmetry breaking~\cite{deForcrand:1999ms,Alexandrou:1999vx,
Engelhardt:2002qs,Bornyakov:2007fz,Hollwieser:2008tq,Hollwieser:2010mj,
Hollwieser:2011uj,Schweigler:2012ae,Hollwieser:2012kb,Hollwieser:2013xja}. This way the vortex model provides a unified picture for the infrared, low
energy sector of QCD explaining both confinement and the chiral and topological
features of the strong interaction. 
However, some of the properties of full QCD are obscured in the P-vortex
(vortex-only) configurations, especially when it comes to topological properties
in connection with fermions. In particular, we address the problem of reproducing a finite chiral condensate in center projected (Z(2)) configurations, using overlap
(and chirally-improved) Dirac operators. Low-lying eigenmodes and
also zero modes are not found in these configurations; the spectra show a
large eigenvalue gap for vortex-only configurations. In~\cite{Hollwieser:2008tq}
the reason for the large gap in the vortex-only case was shown to be connected
to the lack of smoothness of center-projected lattices, {\it i.e.} maximally
non-trivial plaquettes - the vortex plaquettes. In that case the exact symmetry
of the overlap operator is strongly field-dependent, and does not really
approximate the chiral symmetry of the continuum theory. 

In the present work, we want to introduce a new smearing method, in order to
embed the corresponding physics from Z(2) vortex configurations in full SU(2)
configurations by smoothing the thin vortices. We speculate that the infrared
aspects of the QCD vacuum can be understood in terms of thick center vortices,
which can be derived from thin vortex structures.

\section{Method}\label{sec:meth}
\begin{figure}[h]
	\centering
	\includegraphics[width=1.0\linewidth]{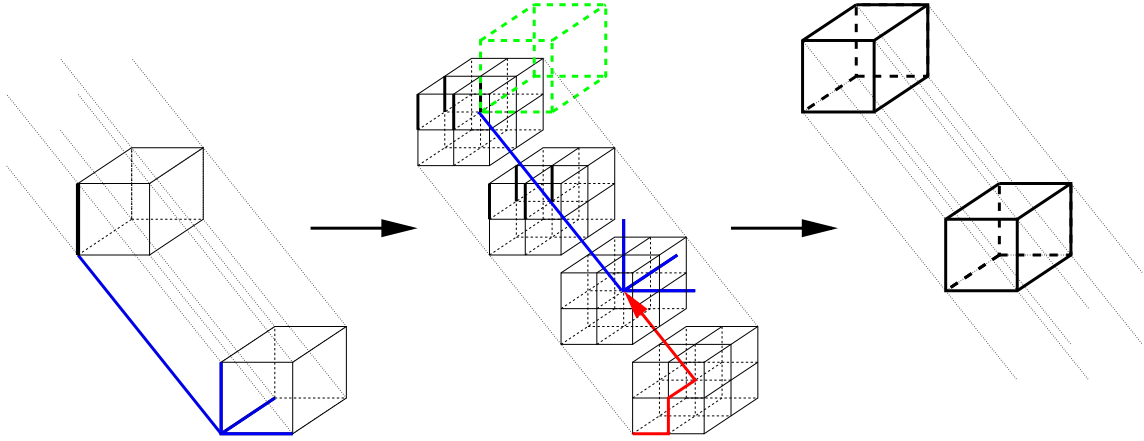}
	\caption{{\bf Refinement routine}: Example of one $-\mathbbm{1}$ (fat z-) link in
	the upper cube of the left lattice part giving eight $-\mathbbm{1}$ and
	eight $\mathbbm{1}$ links in the refined (central) lattice part.
		{\bf Smeared blocking procedure}: After refinement and smearing the
	blocking is not performed starting at (1,1,1,1), the lower, left corner, but
	half an original ({\it i.e.} one refined) lattice spacing forward in every
	space-time direction, {\it i.e.} (2,2,2,2) on the refined lattice, indicated
	by the red arrow (the time direction is indicated by the fine lines
	connecting the space-like cubes). Without vortex smearing between steps 2
	and 3 the lattices before and after the whole procedure are actually the
	same; with vortex smearing the re-blocked lattice gives a smeared version of the original lattice.}
	\label{fig:refblk}
\end{figure}

The idea is to smooth out the thin vortices to regain a finite thickness, {\it
i.e.} we distribute the center vortex flux of the vortex plaquettes, {\it i.e.}
Tr $U_{\mu\nu}=-1$, to several (neighboring) plaquettes. This thickens the
vortices in the sense that the center flux is not restricted to a singular
surface but spread out over a few lattice spacings. On the original lattice,
vortex structures are only a few lattice spacings apart and smearing leads to
distortions of the vortex structure. Hence, we  put the vortex configuration on
a finer lattice. For Z(2) gauge links, the refinement procedure can be defined straightforwardly and yields exactly the same vortex structure but on a finer lattice. 
The refinement procedure is illustrated in Fig.~\ref{fig:refblk}. We double the
number of links in each direction, hence the lattice volume increases by a
factor of $2^4=16$. If the initial link was $\mathbbm{1}$, we only insert
two $\mathbbm{1}$ links; however, if the initial link has value $-\mathbbm{1}$, we
insert a $\mathbbm{1}$ and a $-\mathbbm{1}$ link in forward direction. The new
link pairs are copied forward by half the initial lattice spacing in all orthogonal
directions, {\it e.g.} an x-link $U_x(\vec x)=-\mathbbm{1}$ at $\vec
x=(x,y,z,t)$ gives $\tilde U_x=\mathbbm{1}$ at $(x,y,z,t)$, $(x,y+1/2,z,t)$,
$(x,y,z+1/2,t)$, $\ldots$, $(x,y+1/2,z+1/2,t+1/2)$ and $\tilde U_x=-\mathbbm{1}$ at $(x+1/2,y,z,t)$, $(x+1/2,y+1/2,z,t)$, $(x+1/2,y,z+1/2,t)$, $\ldots$, $(x+1/2,y+1/2,z+1/2,t+1/2)$. Just as refinement gives the same vortex structure on the finer lattice, the
inverse procedure, blocking, again gives the original configuration. During
blocking, the copies between the coarse lattice planes are thrown away and the two refined links, {\it e.g.}
$\tilde U_x(\vec x=(x,y,z,t))$ and $\tilde U_x(\vec x=(x+1/2,y,z,t))$, are multiplied
to reproduce the original $U_x(\vec
x=(x,y,z,t))=\mathbbm{1}\cdot\pm\mathbbm{1}=\pm\mathbbm{1}$ link. On the refined
lattice, however, one now has the advantage that deformations of the vortex surface within the original lattice
spacing still yield the correct vortex structure after blocking. 

\begin{figure}[h]
	\centering
	\includegraphics[width=.8\linewidth]{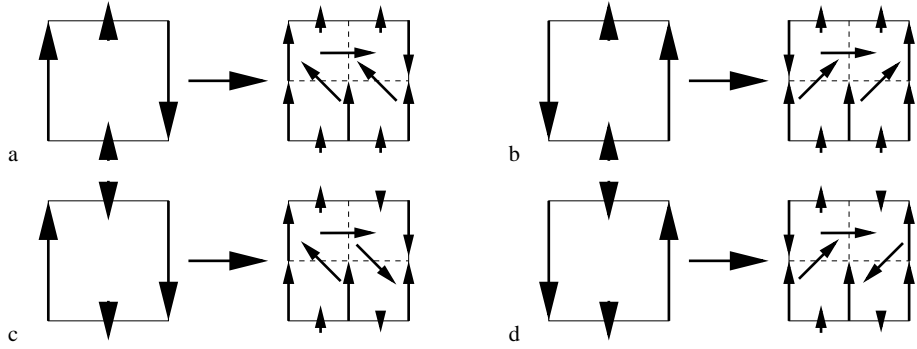}
	\caption{Examples of link configurations giving uniform center vortex flux
		distribution among the four refined plaquettes corresponding to the initial center
		vortex plaquette. The individual link rotations are given by $+\pi/2$,
		$\pm\pi/4$ and $\pm3\pi/4$. Each refined plaquette carries one fourth of the
	initial center vortex flux; a),c) $\exp{i\pi}=-\mathbbm{1}$ and b),d)
$\exp{-i\pi}=-\mathbbm{1}$. In order to minimize the orthogonal plaquettes (the
ones orthogonal to the displayed plane) we apply "2D gauge transformations" (see
text) at the central points of the plaquettes.} 
	\label{fig:reflux}
\end{figure}

The goal is to distribute the center vortex flux $-\mathbbm{1}$ symmetrically
among the refined plaquettes making up the original vortex plaquette. Therefore we smear
all four (refined) links within the original vortex plaquette by individual link
rotations. In Fig.~\ref{fig:reflux} we show examples of link configurations to distribute the center vortex flux $\exp{i\pi}=\exp{-i\pi}=-\mathbbm{1}$ uniformly among the
four refined plaquettes, each carrying one fourth of the initial center vortex
flux. The uniform distribution is of course only guaranteed if we apply all link
rotations of $\pm\pi/4$ and $\pi/2$ in the same U(1) subgroup. 

Depending on the vortex structure, the plotted link configurations may still
cause maximally non-trivial plaquettes in directions orthogonal to the displayed
plane. Hence we apply a gauge rotation to the four smeared links, multiplying
them with a random SU(2) vector at the central point of the original vortex
plaquette. Using this "2D gauge transformation" we can minimize the affected plaquettes
orthogonal to the original or refined vortex plaquette. This way we eliminate
maximally non-trivial plaquettes and get smooth vortex configurations on refined
lattices. We finally block these smeared lattices again in order to get
smooth vortex configurations on the original lattice.

In terms of Z(2) lattices, we have seen that blocking is the exact inverse
procedure to refinement. Hence, blocking a refined lattice exactly gives us
the same links and plaquettes present in the original lattice, and therefore also the
same vortex structure. Since our smearing routine only changes links at half the
initial lattice spacings, {\it i.e.} links dividing the original plaquettes into
four refined plaquettes, blocking trivially (the mentioned links are thrown
away) restores the initial Z(2) link configuration and therefore the original
vortex structure again. However, if we do not block the refined lattice starting
at original lattice sites, but at new, refined lattice sites, more precisely one
refined  (half an original) lattice spacing away from original lattice sites in
every space-time direction, as indicated in Fig.~\ref{fig:refblk} by the red
arrow, we have to multiply several smeared SU(2) links instead of
$\pm\mathbbm{1}$s and end up with a SU(2) instead of a Z(2) configuration. This
new, blocked configuration now represents a smeared version of the original
Z(2) lattice, since the smeared links are derived from the original lattice
after refinement. This procedure we will call "smeared blocking" in the
following, but we should note that the vortex structure extracted from the
smeared blocked lattices is not exactly the same as the original vortex
configuration.

\section{Results}\label{sec:res}
In order to test our method we use 1000 Z(2) vortex configurations on $16^4$
lattices, derived from thermalized Luescher-Weisz SU(2) gauge field
configurations on $8^4$ lattices at coupling $\beta=3.3$
($\sigma_{lat}=0.1112\pm0.0017$, $a=0.1495\pm0.0012$fm) after direct maximal
center gauge (MCG), projection and Z(2) refinement. 
Fermion eigenmodes are calculated using the MILC~\cite{milc} code.

\begin{figure}[h]
	\centering
	a)\includegraphics[width=.51\linewidth]{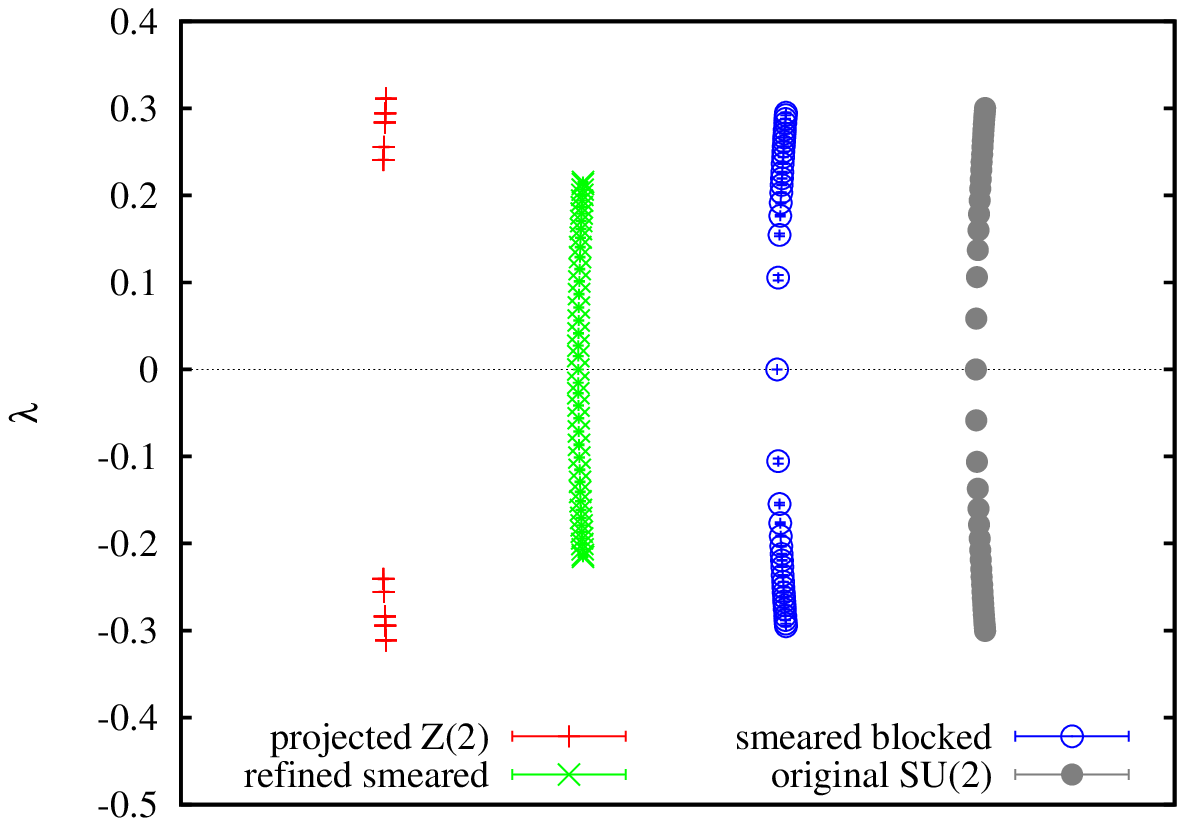}b)\includegraphics[width=.45\linewidth]{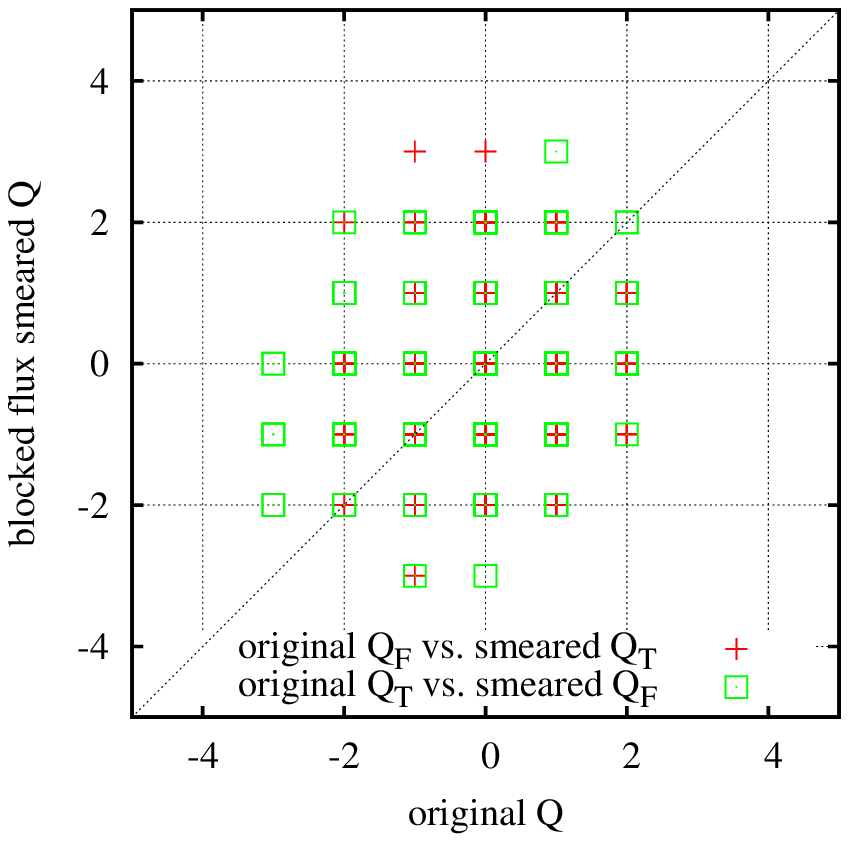}
	\caption{a) 20 lowest overlap eigenvalues for original (full) SU(2), Maximal Center Gauge projected Z(2) and vortex smeared configurations. b) Scatter plot of fermionic and gluonic topological charge correlations between original (full) vs. blocked vortex flux distribution smeared configurations.}
	\label{fig:evs}
\end{figure}

In Fig.\ \ref{fig:evs}a) we display the twenty lowest-lying complex
conjugate eigenvalue pairs of the overlap Dirac operator~ \cite{Narayanan:1994gw}, for center-projected, vortex smeared and original configurations. For the spectra on the refined lattices, the eigenvalues are multiplied by a factor two,
to account for the refinement effect. The refined smearing completely removes
the eigenvalue gap of the center projected configurations and reproduces a
finite number of (near-) zero modes. The spectrum for configurations after the
smeared blocking procedure approaches the original result very well. 

If we want to recover the topological structure of the original (full) SU(2)
configurations from the vortex smeared configurations, we face ambiguities: The
gluonic topological charge $Q_T$, given by the integral of the gluonic charge
density $q(x) = \frac{1}{16\pi^2} \, \mathrm{tr} (\mathcal F_{\mu\nu}
\tilde{\mathcal F}_{\mu\nu})$, is measured after cooling or standard ({\it e.g.}
LOG) smearing, which destroys the relevant vortex structures. The fermionic topological
charge $Q_F = \mbox{Tr} (\gamma_5 D_{ov}) = n_- - n_+ = \text{ind} D_{ov}$ is
given by the difference of right- and left-handed Dirac zero modes, but it is
not necessarily correlated to the vortex topological
charge~\cite{Engelhardt:2002qs}, since the vortex configurations do not
represent a topological torus, as there are monopoles and Dirac strings present.
During cooling or (standard) smearing, monopoles and Dirac strings are smoothed out or fall
through the lattice and the toroidal topology is restored, hence $F\tilde F$
approaches the index topological charge. However, the vortex finding property
gets lost and the vortex topological charge quickly vanishes~\cite{Hollwieser:2012kb}.
In view of the above concerns, it is not surprising that the correlation of
either fermionic $Q_F$ or gluonic topological charge $Q_T$ of the vortex smeared
configurations with the original topological charge is not very good, see
Fig.~\ref{fig:evs}b). The vortex topological charge of original and smeared
configurations is essentially the same, since we deal with identical vortex
structures, however, vortex topological charge depends on the orientation of the (thick) vortex surfaces. The (thin) Z(2) vortices lack any information of orientation and in
order to calculate the vortex topological charge, orientation is applied randomly
to the vortex surfaces. Similarly, during vortex smearing, by replacing Z(2)
gauge links with rotations in the SU(2) space, we automatically give the vortex
surfaces a random orientation in color space, which influences the gluonic
topological charge. Since these two procedures are independent, we can not expect that the smeared vortex configurations or the vortex topological charge in general give comparable results for individual configurations. 
However, in~\cite{Bertle:2001xd} it was shown that the vortex topological charge gives a good estimate of the
topological susceptibility of the gauge field ensemble. The concepts of blocking
and smoothing during vortex topological charge calculation were also discussed
in~\cite{Bertle:2001xd}; the latter reduces vortex topological charge and
susceptibility, since it removes short range fluctuations of the vortex
structure. On our original $8^4$ (refined $16^4$) lattices with original lattice spacing
$a\approx0.15$fm, 1-2 (2-3) blocking steps seem appropriate during the vortex
topological charge calculation in order to arrive within the range of a physical vortex thickness of $~0.4$fm~\cite{DelDebbio:1996mh}.

\begin{figure}[h]
	\centering
	a)\includegraphics[width=.48\linewidth]{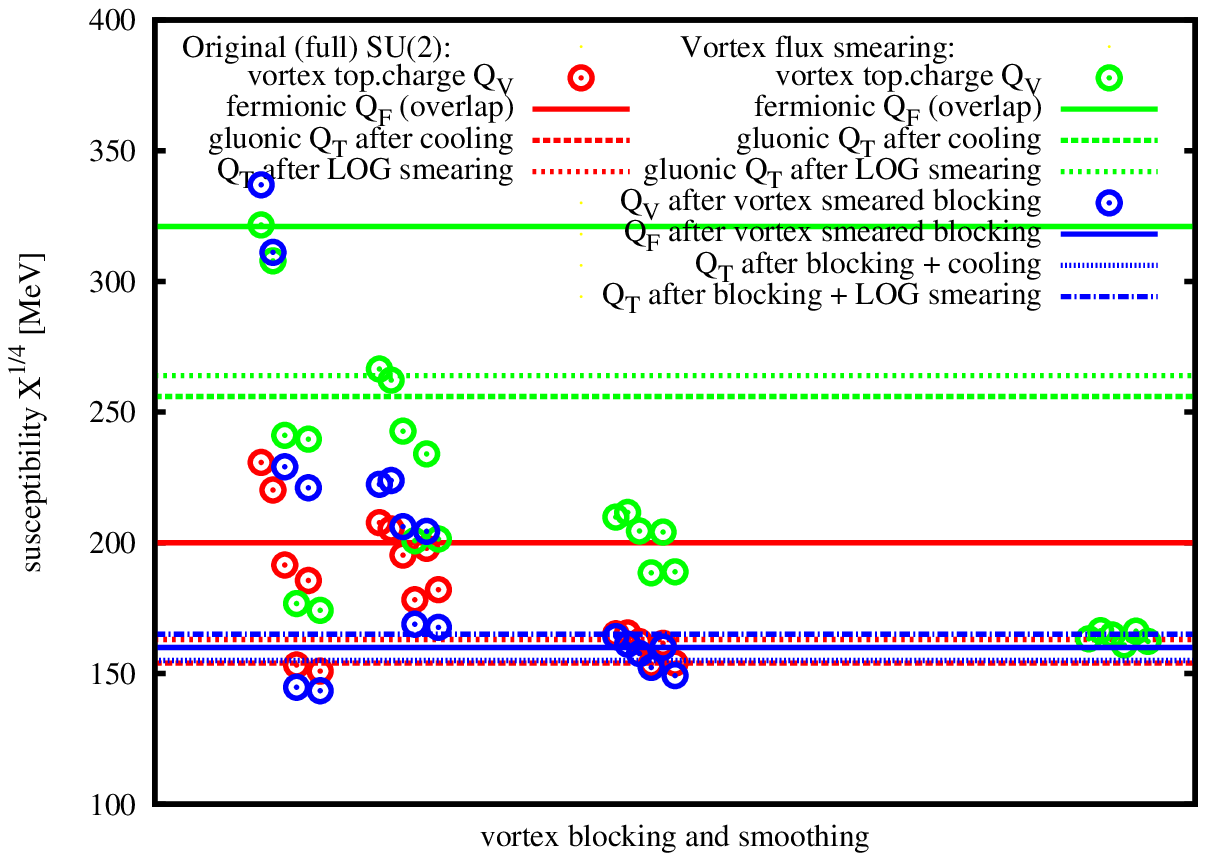}b)\includegraphics[width=.48\linewidth]{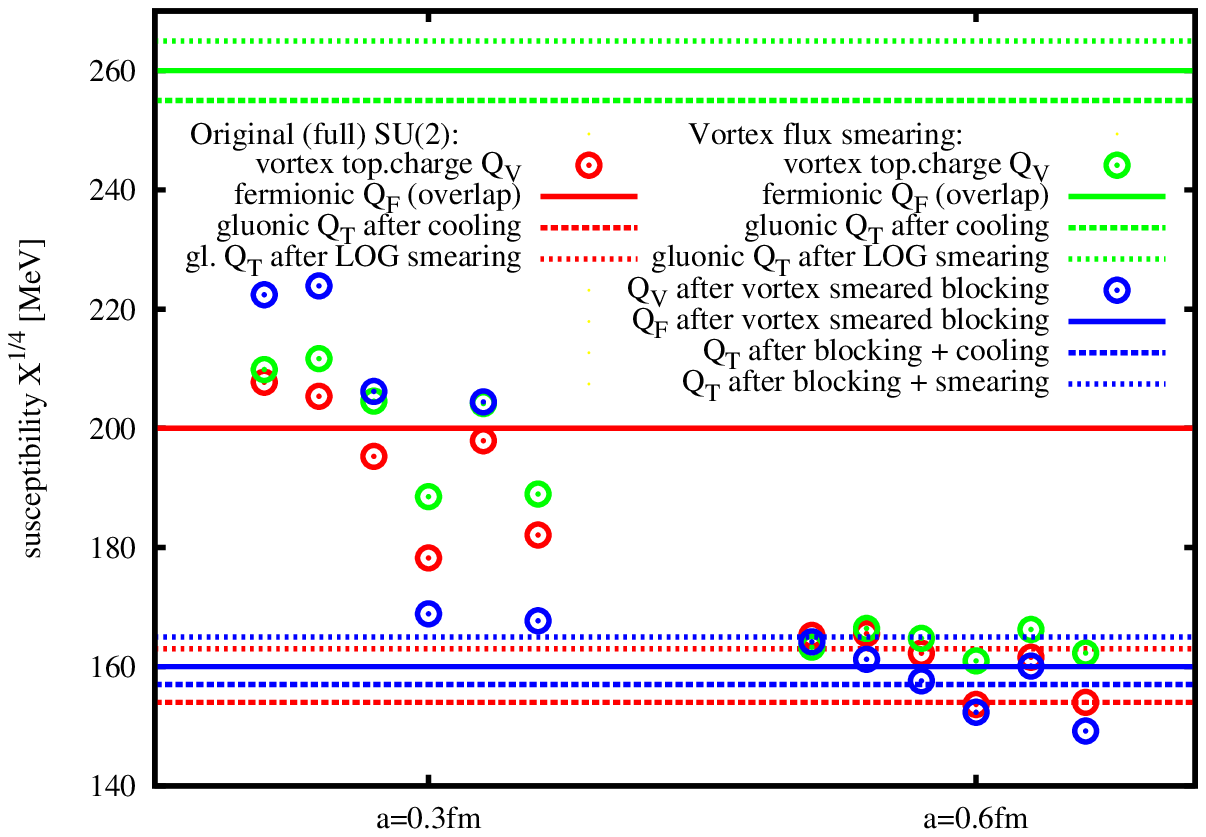}
	\caption{Topological susceptibility from fermionic $Q_F$, gluonic $Q_T$ and
	vortex topological charge $Q_V$ for original (full) SU(2) and vortex flux
	distribution smeared configurations. In a) we show all blocking and
	smoothing steps for $Q_V$ (vertical groups of points correspond to 0,1,2 and
	3 blocking steps, the horizontal splitting within the group indicates
	0,1,2,3,4 or 5 smoothing steps, for details see~\cite{Bertle:2001xd}), in b) we show 1-2 blocking steps for $8^4$ lattices, {\it i.e.} original and smeared blocked configurations, and 2-3 blocking steps for refined ($16^4$) lattices, resulting in lattice spacings $a\approx0.3-0.6$fm. }\label{fig:susc}
\end{figure}

In Fig.~\ref{fig:susc} we show the topological susceptibilities for
original (full) SU(2) and smeared configurations. 
The first thing we note is that for our original SU(2) gauge ensemble,
the topological susceptibilities from fermionic and gluonic topological charge
definitions are not consistent, $\langle Q_F^2\rangle/V=(200$MeV$)^4$ and $\langle
Q_T^2\rangle/V=(160$MeV$)^4$ (averaging cooling and LOG smearing $Q_T$), presumably caused
by our small original lattice volume of about $(1.2$fm$)^4$. It should be noted, however, that the vortex topological susceptibility reproduces these
values with one, respectively two blocking steps, averaging over the
corresponding smoothing steps. Next, we see that for the refined smeared
configurations, the gluonic and vortex topological charges lead to much higher
susceptibilities, caused by the artificial vortex fluctuations introduced during
the refined smearing process. After blocking, however, the original
results are reproduced, shown in Fig.~\ref{fig:susc}b). The gluonic topological
susceptibilities after cooling and (LOG) smearing the vortex smeared
configurations agree with the original (full) SU(2) values
and vortex topological charge also matches the original averages. Concerning
fermionic topological susceptibility, the results are consistent with
gluonic and vortex topological susceptibilities. The refined version gives
$(260$MeV$)^4$, after blocking we find $(160$MeV$)^4$. The results show that vortices are
indeed able to reproduce the topological susceptibility of full Yang-Mills
theory, either via vortex topological charge or gluonic and fermionic definitions after vortex smearing.

\section{Conclusions and Outlook}
We presented a method to smear Z(2) vortex configurations such as to embed
vortex physics into a full SU(2) gauge configuration framework. The main goal
was to remove the eigenvalue gap observed for overlap fermions in
center-projected Z(2) vortex configurations. In order to maintain the original
vortex structure we have to put the Z(2) configurations on finer lattices, where
we can distribute the center vortex flux of the vortex plaquettes onto the refined plaquettes making up the original vortex plaquette. 
Our method thickens the vortices in the sense that the center flux is
not restricted to a singular surface but spread out over a few lattice spacings.
Besides refined smearing, we also discuss a method to block the smeared lattice
back to its original size. With the various methods we can also reproduce
topological properties of the original gauge fields; on vortex smeared lattices,
the different definitions of topological charge, {\it i.e.} fermionic, gluonic
and vortex topological charge, result in comparable susceptibilities. However,
one-to-one correlations of topological charge for individual configurations are
not observed. The reason is that gluonic topological charge definitions are
usually applied after cooling or standard smearing, both transforming Monte Carlo
configurations into smooth gauge fields without or with strongly distorted
center vortex excitations. Thin center vortex gauge fields, {\it i.e.} Z(2)
configurations, on the other hand lack any information on the orientation of
thick center vortices, which is crucial for topological charge determination.
During vortex smearing or vortex topological charge determination, we apply
random orientations to the vortex sheets and cannot expect to reproduce the
original topological charge. However, earlier results and the analysis presented
here and in more detail in~\cite{Hollwieser:2015koa},
show that vortex gauge fields reproduce the original topological charge susceptibility via vortex topological charge definition and via fermionic or gluonic definitions after the introduced vortex smearing methods.

\bibliographystyle{utphys}
\bibliography{latt14.bbl}

\end{document}